\documentstyle[12pt,psfig,subfig]{article}
\begin{document}
\bibliographystyle{unsrt}
\begin{center}
{\Large{\bf Vibrational dynamics of solid poly(ethylene oxide)}} \\
\vspace{2cm}
{\large{\bf M. Krishnan\footnote{email:mkrishna@jncasr.ac.in}, and S. Balasubramanian\footnote{Corresponding Author. email:bala@jncasr.ac.in}}}  \\
\vspace{0.5cm}
Chemistry and Physics of Materials Unit, \\
Jawaharlal Nehru Centre for Advanced Scientific Research,\\
Jakkur, Bangalore 560064, India.\\
\end{center}
\begin{center}
{\large Abstract}
\end{center}
Molecular dynamics (MD) simulations of crystalline poly(ethylene oxide) (PEO) have been carried out in 
order to study its vibrational properties.
The vibrational density of states has been calculated using a normal mode 
analysis (NMA) and also through the velocity autocorrelation function of the atoms.
Results agree well with experimental spectroscopic data.
System size effects in the crystalline state, studied  through a comparison between
results for 16 unit cells and that for one unit cell has shown important differences 
in the features below 100~cm$^{-1}$.
Effects of interchain interactions are examined by a comparison of the 
spectra in the condensed state to that obtained for an isolated oligomer
of ethylene oxide.
Calculations of the local character of the modes indicate the presence of collective
excitations for frequencies lower than 100~cm$^{-1}$, in which around 8 to 
12 successive atoms of the polymer backbone participate. The 
 backbone twisting of helical chains about their long axes is dominant
in these low frequency modes. 

\vspace*{0.5cm}
\newpage    

\par
\section{Introduction}
\par
Solid polymer electrolytes (SPE), composed of inorganic salts solvated in solid, high
molecular weight polymer matrices, have been the focus of intense theoretical and
experimental research because of their
applications as solid state batteries~\cite{halley_power,wakihara}. 
In spite of their wide technological applications, the precise mechanism of 
conduction in these materials is still unclear and remains a pursuit of
interest. In non-polymeric crystalline and glassy electrolytes,
the ionic species hop from one site to another within a rigid host frame.
However, the conduction mechanism in solid polymer electrolytes is
believed to be different~\cite{brucebook,bruce1,angell_jpcm}. NMR studies of 
line shape and relaxation rates in
poly(ethylene oxide)-lithium salt complexes have demonstrated a relationship between
the motion of the Li$^+$ ions
and the segmental motion of the PEO chains~\cite{donoso}. 
This notion gains support from other experimental
studies such as Field-Gradient NMR spin echo technique~\cite{boden}, 
 and Quasi-elastic Neutron Scattering (QENS)~\cite{qens}. 
SPE, in
general, contain both amorphous and crystalline regions, with conduction
being facile in the amorphous regions. This is probably        
due to differences in the chain dynamics~\cite{angell}.
The work of Armand and coworkers has shown that ion mobility in solid
electrolytes is of a continuous, diffusive type, 
in the amorphous regions~\cite{berthier}. 
NMR, differential scanning calorimetry and electrical conductivity 
studies have demonstrated that both cations and anions are mobile in the
amorphous phase~\cite{gorecki}. 
Recently, Bruce and coworkers have shown
that a SPE with a ratio of ether oxygen to lithium of 
6:1, exhibits higher conductivity in its crystalline phase as
compared to its amorphous phase and that the ion transport is dominated by
the cations~\cite{bruce2}. 
In such complexes,          
cross-linking between a pair of polymer chains 
results in a channel like structure. The ion moves in this channel,
aided by the
segmental motion of the polymer chain pair.

A large number of simulations have been carried out on PEO and PEO-salt
complexes, both in their crystalline and amorphous phases over the last decade.
Molecular dynamics studies by de Leeuw et al, using an united atom model (UAM) for
PEO chains in their molten state, have examined the motion of methylene groups
which they term as segmental motion~\cite{leeuw}. 
Neyertz et al have performed extensive molecular dynamics simulations of bulk 
crystalline PEO~\cite{sylvie-94,thesis}, PEO melts~\cite{neyertzmelt}, crystalline NaI-PEO~\cite{sylvie-acta95}
and amorphous NaI-PEO~\cite{sylvie-cps95} systems. The simulations of Muller-Plathe and coworkers
on PEO-LiI complexes~\cite{florian-95} have demonstrated that the ether oxygens belonging to consecutive 
monomers of a PEO chain coordinate to the same $Li^{+}$ ion and that this segmental coordination is 
argued to be the driving force for salt dissolution. 
Laasonen and Klein have performed MD simulations of both crystalline and amorphous PEO-NaI
complexes~\cite{laasonen}. Their study showed that ion pairing is most
probable in the crystalline rather than in the amorphous phase. 
Halley and coworkers
have studied the structure of amorphous PEO using the Parrinello-Rahman 
method recently~\cite{halley2001}. 
MD simulations have also been employed by 
Smith and coworkers to 
elucidate the structure and dynamics of  poly(propylene oxide) 
melts~\cite{smith_peo},  PEO-salt complexes~\cite{smith} and aqueous 
PEO solutions~\cite{smith_h2o}, using a potential model derived from 
{\em ab initio} calculations~\cite{smith_spectrochim}.

It is our endeavor here to characterize the vibrational modes of the polymer backbone
in the pristine polymer, with an emphasis on the low frequency modes. 
This could enhance our understanding on the exact relationship between the 
dynamics of the  polymer and that of the cation in the PEO-salt complexes. 
We carry out this study with the
thought that understanding the evolution of the vibrational modes from the crystalline
phase of pure PEO to its nature in the glassy state without and with the salt is 
crucial in mapping the precise mechanism of ion transport in these systems~\cite{carini,mustarelli}. 
Normal
mode analyses to characterize vibrational spectra have been employed successfully
to describe the dynamics of glassy systems~\cite{elliott_papers}, and are likely
to be employed in the study of jammed granular
materials~\cite{nagel}. These calculations have also helped our understanding of vibrational
modes of polyethylene~\cite{hess_avg}, of proteins in solution~\cite{karplus}, and in 
determining the flexibility of proteins, and in the thermodynamics of hydration water
in protein solutions~\cite{verma_jpcb}.
Unlike polyethylene, PEO adopts a distorted
helical conformation in its crystalline state, with the distortion arising from strong
intermolecular interactions. It is thus important that such normal mode calculations 
be performed for the crystalline state, rather than for an isolated oligomer.
In this paper, we limit
ourselves to the study of the vibrational spectrum of PEO in its crystalline state.
Analyses of these modes under other state and phase  
conditions will be examined in future. In anticipation of our results, 
we have identified the existence of segments in the polymer backbone consisting 
of around 8 to 12 atoms, to exhibit significant atomic displacements, for vibrational modes
with frequencies up to around 100~cm$^{-1}$,  and that the modes with frequencies 
below 60~cm$^{-1}$ involve twisting of the skeletal backbone.
The paper is divided as follows. Details of the simulation and the methods of 
analyses are presented in the next section. Later, we discuss the 
results obtained from our study. Details on obtaining the elements of the dynamical matrix 
are provided in the Appendix.

\par
\section{Details of Simulation}
\par

The unit cell
of crystalline PEO is monoclinic with the PEO chains in a (7/2)
distorted helical conformation with TTG sequence~\cite{takahashi}. An unitcell consists of four PEO
chains with 21 backbone atoms in each chain. The MD runs reported here 
were initiated from configurations generated from this crystal structure. 
The two ends of a polymer chain were assumed to be bonded across the simulation cell 
boundary~\cite{sylvie-94,laasonen} to eliminate end effects. 
MD simulations were performed primarily for a system that we
describe as $\langle422\rangle$, which contained 4 unitcells along the a-axis, 2 unitcells along 
the b-axis and 2 unitcells along the c-axis, containing a total of 3136 atoms.
To study system size effects and effects of inter-chain interactions, we have 
performed additional simulations of only one unitcell, and also 
of one isolated finite length polymer chain. Thus the simulations for the $\langle422\rangle$ cells and 
that for the one unitcell contained chains that had no ends, while that for the isolated
molecule,  contained a chain with two ends.
An all atom model (AAM), with explicit 
consideration of hydrogen atoms was used, to properly account for the 
steric interactions expected to be dominant in the crystalline state. The simulations
were carried out in the canonical ensemble at 5K for the NMA and in the 
constant pressure ensemble at 300K and 1~atmosphere to test 
the stability of the simulated crystal.
Temperature control was achieved
by the use of Nose-Hoover chain thermostats~\cite{martyna}, using the PINY-MD 
program~\cite{piny-md}. Long range interactions were treated using the Ewald 
method with an $\alpha$ value of 0.3~\AA$^{-1}$, and 2399 reciprocal space 
points were included in the Ewald sum~\cite{hansen}.
The methylene groups were treated as rigid entities, enabling us to employ 
a timestep of 1~fs.
To obtain the vibrational density of states,
we performed these calculations at a temperature of 5K, where the atoms could be 
expected to be near their equilibrium positions. 
The equilibration period for the single molecule, the unitcell, and the $\langle422\rangle$ system were
65~ps, 100~ps, and 750~ps, respectively. These were followed by an analysis run of duration 30~ps
during which the coordinates and velocities
of each particle were stored at regular intervals.
Velocities were dumped
every time step to obtain the power spectrum of their time correlation 
functions and the coordinates were dumped every 10 fs.   

The simulations were performed with
a force field 
obtained from the work of Neyertz~\cite{sylvie-94} with the 
torsional parameters of the CHARMm model~\cite{charmm}.
The potential parameters are given in Table 1. The non-bonded interactions were
truncated at 12\AA~for the $\langle422\rangle$ crystal, at 3.25\AA~for one unitcell, and at 12\AA~for
 the single molecule runs.
The lower cutoff value for the unitcell run is necessitated by the fact that the
 interaction cutoff should be
less than half of the minimum distance between any two opposite faces
of the simulation cell. 

The potential energy of the
system can be expanded in terms of atomic displacements from an equilibrium configuration as,
\begin{equation}
U\left(q_{1},q_{2},...,q_{n}\right) = U\left(q_{01},q_{02},...,q_{0n}\right) + 
\left(\frac{\partial U}{\partial q_{i}}\right)_{0} \eta_{i} +
 \frac{1}{2}\left(\frac{\partial^{2} U}{\partial q_{i} \partial q_{j}}\right)_{0} 
\eta_{i} \eta_{j} + ...
\end{equation}
where the subscript 0 represents the equilibrium configuration and $\eta_{i}$
are the deviations of the coordinates from equilibrium, represented as,
\begin{equation}
q_{i} = q_{0i} + \eta_{i}
\end{equation}
The elements of the Hessian matrix are given by
\begin{equation}
H^{\alpha\beta}_{ij} = \frac{1}{\sqrt{m_{i}m_{j}}} \left(\frac{\partial^{2}U}{\partial \beta_{j} 
\partial \alpha_{i}}\right)
\end{equation}
where i,j represent particle indices and $\alpha$, $\beta$ represent the spatial
coordinates x,y,z. $m_{i}$ is the mass of particle i. 
A simple scheme to obtain some of these Hessian elements efficiently is 
provided in the Appendix. {\em All 
Hessian elements obtained from such analytical expressions were checked against
numerical second derivatives~\cite{abramovitz} within our normal mode analysis code, and were 
found to match.}
The eigenvalues and eigenvectors of the Hessian matrix were examined to 
understand the vibrational dynamics of PEO. 
The frequency, $\nu_{s}$ ,of a particular mode of vibration, s,
is related to its eigen value, $\lambda_{s}$ , by  
\begin{equation}
\lambda_{s} = \left(2\pi\nu_{s}\right)^{2}
\end{equation}

With these set of interactions, the initial pressure of the $\langle422\rangle$ system was
found to be around 4000 atmospheres. Hence, we
performed a MD run in the NPT ensemble for 200 ps under ambient conditions. The change
in volume was found to be 1.3\% from that of the experimental crystal.
Time correlation functions of atomic velocities were calculated, 
and were Fourier transformed to obtain the power spectra. 
These were compared with the spectrum obtained from the NMA. 
The spectra were convoluted with a Gaussian function of width 4cm$^{-1}$ so as to 
provide a width to the spectral features. 
The helical axis of a PEO chain possesses a 7-fold rotational
symmetry. In addition, 7 C$_2$ axes lie perpendicular to it, making the molecular symmetry of
PEO to be D$_7$~\cite{takahashi}. We have characterized the symmetry of the normal 
modes by studying the transformation of atomic displacements,
on application of the symmetry operations for this group. Specifically,
the operation by the C$_2$ axes enables a distinction between the three irreducible 
representations, A$_1$, A$_2$, and E. 

We have also characterized the spatial extent of the modes of vibration using a quantity called the 
local character, defined as~\cite{karplus,elliott}
\begin{equation}
Local character\left(j\right) = \sum_{i=1}^{3N}u_{ij}^{4}
\end{equation}
where $u_{ij}$ is the i$^{th}$ component of the j$^{th}$ eigenvector.
The local character value can range from 0 to 1 and it determines 
the extent of localization of a particular mode.

To obtain quantitative information on the length of the segment involved in the 
low frequency modes, we have defined a
quantity called the Continuous Segment Size (CSS). We calculate the displacement of
the $k^{th}$ atom due to a $i^{th}$ mode using the expression 
\begin{equation}
<\delta r^{2}_{ik}> = k_{\rm B}T \frac{\bracevert \vec{u}_{i}^{k}\bracevert^{2}}{m_{k}\omega_{i}^{2}}
\end{equation}
where $m_{k}$ is mass of $k^{th}$ atom, $\vec{u}_{i}^{k}$ is
 the vector formed by the components of the i$^{th}$
 eigen vector contributed by the k$^{th}$ atom, $T$ is temperature, and $\omega_{i}$ is the 
frequency of the i$^{th}$ normal mode. 
We then check if the displacement is greater than a specified cutoff.
If $n$ successive backbone atoms of a chain each have displacements greater
than the displacement cutoff, they are defined to constitute a segment with CSS = $n$. Similarly,
 CSS was calculated for all possible modes with
different vibrational frequencies from which the average segment size for a given
frequency was calculated.  

\section{Results and Discussion}
For crystalline systems, a close
match between the experimentally determined cell parameters and that obtained from simulations 
is a crucial first step in the veracity of the parameters used. 
We show in Figure~1, the time 
evolution of the cell parameters at 300K and 1~atmospheres, generated by a MD run
in the constant pressure ensemble using the Parrinello-Rahman method~\cite{parr-rahman}.
The $b$ and the $\beta$ parameters of the unit cell exhibit a relaxation from the 
zero time experimental value, due possibly to relatively minor deficiencies in the 
interaction model used to represent this system. 
The constancy of the cell parameters for over 100~ps, shows that the simulated crystal is in 
a stable state, and that the potential parameters are indeed able to reproduce
the high demands of the crystal symmetry.
Table 2 compares the cell parameters obtained from our simulations to the experimental data.

An interesting feature associated with polymer dynamics is the variation of the
various vibrational modes of a chain encountered during its transformation from
the isolated state to the crystalline state. The vibrational spectra of a single
molecule, one unitcell and that of the $\langle422\rangle$ crystal are compared in Figure~2a. 
When the chains assemble to form a crystal, each chain might prefer a new conformational
state relative to its structure in the isolated state. A comparison of the vibrational
spectra between that of one isolated molecule of finite length and of the $\langle422\rangle$ system
provides information on the effect of intermolecular interactions. 
As expected, the vibrational states of the isolated molecule showed
marked differences from the crystalline state, particularly in the low frequency regions,
where large amplitude, collective motions are predominant (see later).
 However, there are no significant
changes in the high frequency regions of the spectrum. The spectrum for the one unitcell
compares well with that of the $\langle422\rangle$ crystal. However, the features in the VDOS of the 
$\langle422\rangle$ crystal is much better resolved, particularly at low frequencies. For instance, 
the features at around 40~cm$^{-1}$ and 
75~cm$^{-1}$ are clearly evident in the larger system than in the spectrum for the unitcell,
pointing to effects of long range interactions. The spectrum for the $\langle422\rangle$ system is shown
in an expanded scale in Figure~2b. The split in the feature below 100~cm$^{-1}$  is evident 
and compares well with experimental infrared (IR) spectra~\cite{rabolt} that shows 
features at 37~cm$^{-1}$, 52~cm$^{-1}$, 81~cm$^{-1}$ and 107~cm$^{-1}$. We do observe a prominent shoulder 
at 108~cm$^{-1}$ which has been attributed to modes involving C-O internal rotation earlier~\cite{rabolt}.
A comparison of the simulated vibrational density of states with the experimental IR and Raman 
spectra, exhibited in Figure~2c, shows that the potential model captures well nearly all the 
vibrational modes~\cite{rabolt,dacosta,branca}. Note that the simulated spectrum is the raw
density of states and does not contain any other terms that are needed to calculate the 
experimental spectra. Thus only the peak positions need to be compared, and not their intensities. 
Almost all the features found in the IR and Raman spectra seem to be present in the simulated
VDOS.
Another noteworthy feature of the spectrum is the
absence of modes with imaginary frequencies which would have corresponded to either the
 presence of atoms away
from equilibrium locations or to a mismatch between the empirical potential function and the 
crystal structure. 

We have also calculated the vibrational spectrum through the
Fourier transformation of the velocity auto correlation function of the atoms.  This
is compared with the VDOS obtained by the NMA in Figure~3.
The agreement between the spectra is excellent except for features above
1200~cm$^{-1}$. Also, the peak at around 1470~cm$^{-1}$ is 
missing in the spectrum obtained through the velocity autocorrelation function (VACF). 
Visualization of the atomic displacements associated with this mode revealed 
HCH bending in methylene groups.
Since all the CH$_{2}$ groups were
constrained to be rigid during the MD runs for the VACF 
analysis, the absence of a peak in the VDOS obtained
from VACF, in this region can be rationalized. The comparison of the VDOS obtained from
the two methods is good, despite the fact that the NMA method is only a harmonic 
approximation to the potential. Close examination of the two shows a marginal (approximately
4-5~cm$^{-1}$) shift to higher frequencies in the spectrum obtained by the NMA method 
relative to that from the VACF.

We visualized the eigenvectors corresponding to different vibrational modes to assign
the nature of atomic displacements 
that are responsible for the spectral features.  In general, our assignments are consistent with earlier
calculations~\cite{yoshihara,lam}.
  In Figure~4a and Figure~4b, we 
display a few of the modes. 
The zero frequency mode characterizes rigid body translation. The features at around 40~cm$^{-1}$,
have earlier been attributed to chain deformations~\cite{rabolt}. Based
on visualization of atomic displacements shown in Figure~4a, we assign these modes specifically to 
the twisting of the polymer backbone. The mode at 88~cm$^{-1}$ arises from torsional
motion around the C-O bond. The A$_1$ mode at 216~cm$^{-1}$ can be assigned to torsions around the 
C-C bond,
while the  510~cm$^{-1}$ feature involves bending of CCO triplets. COC bending is 
observed at 952~cm$^{-1}$ while the  
wagging motion of the methylene groups is found to be present at 1244~cm$^{-1}$, in good 
agreement with IR and Raman measurements~\cite{yoshihara}. The symmetry of the modes are also
shown in Figures~4 agree well with experimental assignments~\cite{yoshihara,rabolt,lam}. 

For 
characterizing the normal modes further, we have calculated the local
character for each mode~\cite{karplus,elliott} which is shown in Figure~5. In the
range, 0 to 160~cm$^{-1}$, the local character value is very small, implying 
the participation of a large number of atoms.
However, the local character does not provide any information 
on the proximity of the atoms that exhibit significant displacement in a mode. Thus
this quantity has to be augmented by a further analysis of the concomitancy or 
proximity of atoms excited in a mode.

We have developed such an index that takes into account the connectivity of the
atoms involved in a mode.
We determine the number of successive atoms, $n$,
that participate in a mode of a given frequency, by calculating  the
distribution of segment sizes, f(n), for that vibrational mode. As a
representative example, the distribution of the CSS for one such vibrational
mode of frequency 44~cm$^{-1}$ is shown in Figure~6 for four selected 
displacement cutoffs. It is our contention that a large number of atoms in 
proximity to each other participate in these modes. This is evident from the 
non-zero value of f(n) for n values in the range of 5 to 10, for some of 
the cutoff values.

The average segment
size was calculated by evaluating the following summation,
\begin{equation}
<CSS> = \frac{\sum_{n=5}^{42} n f\left(n\right)}{\sum_{n=5}^{42} f\left(n\right)}
\end{equation}
In our study, we define a segment as a chain of connected atoms whose displacements
are larger than a specified cutoff, with n $\geq $
5. Values with n $<$ 5, which correspond to intramolecular excitations that 
arise out of bonded interactions like the
stretch, bend and torsion, have been omitted in $<CSS>$ calculations, as their 
origin is trivially known. It should also be noted that the exact values of the average segment 
size will depend on the chosen displacement cutoff. Hence, we have performed
these calculations for a variety of cutoffs and these are exhibited in Figure~7 
as  a function of the frequency of the modes.
Notice that the segment size for the zero frequency modes, {\em i.e.,}
translations, is 42, which is the number of backbone atoms in a given chain.
It is also evident from the figure that the average segment size decays
with increase in frequency.
Figure~7 also shows
the variation of $<$CSS$>$ as a function of frequency, for four different
displacement cutoffs. The displacement cutoff which spans the frequency range
relevant to collective motion, where the the local character value is almost
zero is the one of significance. 
It can be seen that in the frequency range of 10~cm$^{-1}$ and 100~cm$^{-1}$, which
is the range of collective motion, around 8 to 12 successive atoms of the backbone 
are involved in the excitations.

\section{Conclusions}
We have studied the vibrational dynamics of crystalline poly(ethylene oxide)
using molecular dynamics and normal mode analysis. The vibrational density of
states obtained from NMA matches well with that obtained from the Fourier
transformation of velocity autocorrelation function and also with experimental
IR and Raman data~\cite{rabolt,dacosta,branca}. The VDOS of an isolated
PEO chain of finite length  showed  marked differences from that of the crystal
 in the low frequency region where collective modes are
predominant. We have also explored system size effects by comparing the VDOS
obtained from a simulation containing 16 unitcells and that of one unitcell. The spectrum
obtained from the former is much better resolved, particularly at low frequencies,
where three clear features, at 44~cm$^{-1}$, 70~cm$^{-1}$, and 88~cm$^{-1}$ are
observed. 
The results of our calculations agree well with assignments of mode symmetry
by earlier workers, that used standard methods for a single chain of PEO, or for an 
unit cell~\cite{yoshihara,rabolt,lam}.  Such calculations have the added advantage, 
over the MD route presented here, of being able to obtain dispersion of the vibrational modes.
However, the method described here can be used to characterize these vibrations in 
the amorphous and liquid phases of PEO.

The normal modes
obtained from the present analysis were characterized by the local character
indicator and by a new quantity that determines the number of concomitant 
atoms excited 
by a mode, called the CSS. In the range 0 to 160~cm$^{-1}$, the value of 
the local
character indicator was very small, indicating the participation of a large number of
atoms in the vibrational modes in this range. A distribution of segment sizes was
calculated for each mode from which the average continuous segment size was
calculated as a function of the vibrational frequency. The frequency dependence of
$<$CSS$>$ clearly shows collective modes to be present for frequencies less than
100~cm$^{-1}$, in which around 8 to 12 successive
atoms of the backbone participate.  This quantitative analysis is also corroborated 
by visualization of the atomic displacements for the low frequency modes.

The phase behavior of PEO and the evolution of these vibrational modes as 
a function of temperature and their properties in the amorphous phase will form
the objectives of our study in future. 

\section{Acknowledgements}
We thank Prof. S. Vasudevan for enlightening discussions, and Dr. Preston Moore for
the visualization program used in Fig. 4.
\newpage
\par

\section{Appendix 1}
\par
The form of the potential used in our simulations is generic to macromolecular interactions, 
and can be found in Ref.~\cite{sylvie-94}.  Here, we provide only the torsional and 
Coulomb terms.
\begin{equation}
U_{coulomb} = \frac{q_{i}q_{j}}{r_{ij}}
\end{equation}
and
\begin{equation}
U_{torsion} = f\left(cos\phi\right) = \sum^{6}_{i=0} a_{i}cos^{i}\phi
\end{equation}
We provide here a procedure 
to obtain the Hessian elements, which is easy to program. To our knowledge, such a scheme has
not been outlined so far~\cite{case}, and hence we provide it here for pedantic reasons.
We limit these details to contributions from 
the torsional interactions, the reciprocal space part and the real space 
part of the Ewald sum. 
Contributions from other terms in the potential energy are much simpler to 
derive and are not given here.

\subsection{Hessian from the torsional interaction}
The torsional angle $\phi$ between the planes formed by the bond vectors 
$\vec{r}_{12}$, $\vec{r}_{23}$, and $\vec{r}_{34}$ is, 
\begin{equation}
cos\phi  = \hat{A}   \cdot  \hat{B} = A_{\gamma}B_{\gamma}
\label{eq: cosphi}
\end{equation}
where the Einstein summation convention has been used and
\begin{equation}
\hat{A}= \frac{\vec{r}_{12} \times  \vec{r}_{23}}{\bracevert \vec{r}_{12} \times  \vec{r}_{23} \bracevert },~~~and~~~
\hat{B}= \frac{\vec{r}_{23} \times  \vec{r}_{34}}{\bracevert \vec{r}_{23} \times  \vec{r}_{34} \bracevert }
\end{equation}
The second derivative 
of $U_{torsion}$ with respect to a spatial coordinate $\alpha_{n}$, where
$\alpha$ can be either x, y, or z and n=1, 2, ..., N, can be written as,  
\begin{equation}
\frac{\partial ^{2}U_{torsion}} {\partial \beta_{m}\partial \alpha_{n}}=\left(\frac{\partial f} {\partial cos\phi}\right)\left(\frac{\partial ^{2}cos\phi} {\partial \beta_{m}\partial \alpha_{n}}\right) + \left(\frac{\partial cos\phi}{\partial \alpha_{n}}\right)\left(\frac{\partial cos\phi}{\partial \beta_{m}}\right)\left(\frac{\partial ^{2}f} {\partial cos\phi ^{2}}\right)
\label{eq: 2U}
\end{equation}
The derivates of $cos\phi$ can be calculated as follows. 
\begin{equation}
\frac{\partial ^2 cos\phi}{\partial \beta_{m} \partial \alpha_{n}} = A_{\gamma}\frac{\partial ^2 B_{\gamma}}{\partial \beta_{m} \partial \alpha_{n}}  + \frac{\partial B_{\gamma}}{\partial \alpha_{n}}\frac{\partial A_{\gamma}}{\partial \beta_{m}} + B_{\gamma}\frac{\partial ^2 A_{\gamma}}{\partial \beta_{m} \partial \alpha_{n}} + \frac{\partial A_{\gamma}}{\partial \alpha_{n}}\frac{\partial B_{\gamma}}{\partial \beta_{m}}
\label{eq: 2cosphi}
\end{equation}

We can write $A_{\gamma}$ as
\begin{equation}
A_{\gamma} =  \frac{N_{\gamma}^{A}}{D_{A}}
\end{equation}
with
\begin{equation}
N_{\gamma}^{A} =  \epsilon_{\gamma\nu\xi}\left(\vec{r}_{12}\right)_{\nu}\left(\vec{r}_{23}\right)_{\xi}
\end{equation}
and
\begin{equation}
D_{A} = \left[ \sum_{l}\left(N_{l}^{A}\right)^{2}\right]^{\frac{1}{2}}
\end{equation}
where $\gamma$, $\nu$, and $\xi$ stand for any of the three indices x, y, z and $\epsilon_{\gamma\nu\xi}$
is the antisymmetric Levi-Civita tensor of rank 3. A similar expression can be written 
for $B_{\gamma}$.

The first derivative of $A_{\gamma}$ are,
\begin{equation}
\frac{\partial A_{\gamma}}{\partial \alpha_{n}} = 
 \frac{1}{D_{A}}\frac{\partial N_{\gamma}^{A}}{\partial \alpha_{n}} - 
 \frac{N_{\gamma}^{A}}{D_{A}^{3}}\sum_{l} N_{l} \frac{\partial N_{l}^{A}}{\partial \alpha_{n}}
\label{eq: 1A-gamma}
\end{equation}
where,
\begin{equation}
\frac{\partial N_{\gamma}^{A}}{\partial \alpha_{n}} = \sum_{\nu = x,y,z}\epsilon_{\gamma\nu\alpha}\left[\left(\vec{r}_{12}\right)_{\nu}\left(\delta_{n3}-\delta_{n2}\right) - \left(\vec{r}_{23}\right)_{\nu}\left(\delta_{n2}-\delta_{n1}\right)\right]
\end{equation}
The second derivative of $A_{\gamma}$ is calculated as follows.
\begin{equation}
\frac{\partial}{\partial \beta_{m}}\frac{\partial A_{\gamma}}{\partial \alpha_{n}} = 
\frac{\partial \left[\frac{1}{D_{A}}\frac{\partial N_{\gamma}^{A}}{\partial \alpha_{n}}\right]}{\partial \beta_m}
-
\frac{\partial \left[\frac{N_{\gamma}^{A}}{D_{A}^{3}}\sum_{l} N_{l} \frac{\partial N_{l}^{A}}{\partial \alpha_{n}}\right]}
{\partial \beta_m}
\label{eq: 2A-gamma}
\end{equation}
with
\begin{equation}
\frac{\partial \left[\frac{1}{D_{A}}\frac{\partial N_{\gamma}^{A}}{\partial \alpha_{n}}\right]}{\partial \beta_m} =
 =  \frac{1}{D_{A}}\frac{\partial}{\partial \beta_{m}}\frac{\partial N_{\gamma}^{A}}{\partial \alpha_{n}} - \frac{1}{D_{A}^{2}}\frac{\partial N_{\gamma}^{A}}{\partial \alpha_{n}}\frac{\partial D_{A}}{\partial \beta_{m}}
\end{equation}
\begin{eqnarray}
\frac{\partial \left[\frac{N_{\gamma}^{A}}{D_{A}^{3}}\sum_{l} N_{l} \frac{\partial N_{l}^{A}}{\partial \alpha_{n}}\right]}
{\partial \beta_m}
 & = & \frac{N_{\gamma}^{A}}{D_{A}^{3}}
\left[\sum_{l}\left[ N_{l}\frac{\partial ^{2}N_{l}}{\partial \beta_{m} 
\partial \alpha_{n}} + \left(\frac{\partial N_{l}}{\partial \alpha_{n}}\right)
\left(\frac{\partial N_{l}}{\partial \beta_{m}}\right)\right]\right] \nonumber \\
 &  + & \left(\sum_{l}N_{l}\frac{\partial N_{l}}{\partial \alpha_{n}}\right) 
\left(\frac{1}{D_{A}^{3}}\frac{\partial N_{\gamma}}{\partial \beta_{m}} -
 \frac{3}{D_{A}^{4}}N_{\gamma}\frac{\partial D_{A}}{\partial \beta_{m}}\right)
\end{eqnarray}
where,
\begin{equation}
\frac{\partial}{\partial \beta_{m}}\frac{\partial N_{\gamma}^{A}}{\partial \alpha_{n}} = \epsilon_{\gamma\beta\alpha} \left[\left(\delta_{m2}-\delta_{m1}\right)\left(\delta_{n3}-\delta_{n2}\right)- \left(\delta_{m3}-\delta_{m2}\right)\left(\delta_{n2}-\delta_{n1}\right)\right]
\end{equation}
Using equations (\ref{eq: 2cosphi}), (\ref{eq: 1A-gamma}), and (\ref{eq: 2A-gamma}) we can calculate $\frac{\partial^{2}cos\phi}{\partial \beta_{m} \partial \alpha_{n}}$ which 
can be substituted in equation (\ref{eq: 2U}) to get the torsional contribution to the Hessian. 

\subsection{Hessian from the Coulomb interaction}
The reciprocal space energy in the Ewald sum method is,
\begin{equation}
U_{reci} = \frac{1}{V\epsilon_{0}} \sum_{\vec{k} \neq 0}
 \frac{e^{\frac{-k^{2}}{4 \zeta^2}}}{k^{2}}
\left(a_k^2 + b_k^2\right)
\end{equation}
where,
\begin{equation}
a_k = \sum_{j=1}^{N} q_{j} cos\left(\vec{k} \cdot \vec{r}_{j}\right), ~~~~and 
~~~~~~ b_k = \sum_{j=1}^{N} q_{j} sin\left(\vec{k} \cdot \vec{r}_{j}\right)
\end{equation}

Here, $V$ denotes the volume of the simulation cell, $\zeta$ determines the width of the 
Gaussian charge distribution centered on the point charges in the Ewald sum method, $\vec{k}$ denote
the reciprocal lattice vectors, and $\epsilon_0$, the permittivity of free space.

It can be shown that

\begin{eqnarray}
\frac{\partial ^{2}U_{reci}}{\partial \beta_{m} \partial \alpha_{n}}
 &  = & \frac{1}{V\epsilon_{0}}  \sum_{\vec{k} \neq 0}
 \frac{e^{\frac{-k^{2}}{4 \zeta^2}}}{k^{2}}
 \left\{ 2q_{n}k_{\alpha}\left[-k_{\beta}\delta_{mn}\left(a_k cos\left(\vec{k} 
\cdot \vec{r}_{n}\right) + b_k sin\left(\vec{k} \cdot \vec{r}_{n}\right)\right)\right.\right.  \nonumber \\
& + & \left.\left. q_{m}k_{\beta} cos\left(\vec{k} \cdot \left(\vec{r}_{n} - 
\vec{r}_{m}\right)\right)\right] \right\}
\end{eqnarray}

The expression for the real space energy in the Ewald summation is, 
\begin{equation}
U_{real} = \frac{1}{4\pi\epsilon_{0}}\sum_{i=1}^{N}\sum_{j>i}^{N}q_{i}q_{j}\rho
\end{equation}
where,
\begin{equation}
\rho = \frac{erfc\left(\zeta r_{ij}\right)}{r_{ij}}
\end{equation}
The second derivatives of $U_{real}$ are
\begin{equation}
\frac{\partial}{\partial \beta_{m}}\left(\frac{\partial U_{real}}{\partial \alpha_{n}}\right) = \frac{1}{4\pi\epsilon_{0}}\sum_{i=1}^{N}\sum_{j>i}^{N}q_{i}q_{j}\frac{\partial}{\partial \beta_{m}}\left(\frac{\partial \rho}{\partial \alpha_{n}}\right)
\label{eq: 2U-real}
\end{equation}
with
\begin{equation}
\frac{\partial}{\partial \beta_{m}}\left(\frac{\partial \rho}{\partial \alpha_{n}}\right) = \left(\frac{\partial \rho}{\partial r_{ij}}\right) \frac{\partial}{\partial \beta_{m}}\left(\frac{\partial r_{ij}}{\partial \alpha_{n}}\right) + \left(\frac{\partial r_{ij}}{\partial \alpha_{n}}\right) \frac{\partial}{\partial \beta_{m}}\left(\frac{\partial \rho}{\partial r_{ij}}\right) 
\end{equation}

\begin{equation}
\frac{\partial}{\partial \beta_{m}}\left(\frac{\partial r_{ij}}{\partial \alpha_{n}}\right) = \frac{\left(\delta_{jn}-\delta_{in}\right)\left(\delta_{jm}-\delta_{im}\right)}{r_{ij}}\left[\delta_{\alpha \beta}-\frac{\left( \alpha_{j}-\alpha_{i}\right)\left( \beta_{j}-\beta_{i}\right)}{r_{ij}^{2}}\right]
\end{equation}

It can be shown that
\begin{equation}
\frac{\partial \rho}{\partial r_{ij}} = -\left(s_{1} + s_{2}\right)
\end{equation}
and
\begin{equation}
\frac{\partial^{2} \rho}{\partial r_{ij}^{2}} = 2\zeta^{2}s_{1}r_{ij} + \frac{2\left(s_{1} + s_{2}\right)}{r_{ij}}
\label{eq: 2rho-rij}
\end{equation}
with,
\begin{equation}
s_{1} = \frac{2\zeta}{\sqrt \pi} \frac{e^{-\zeta^{2}r_{ij}^{2}}}{r_{ij}}
~~~~~~and ~~~~~~~ s_{2} = \frac{erfc\left(\zeta r_{ij}\right)}{r_{ij}^{2}}
\end{equation}

These can be used in Eq.~\ref{eq: 2U-real} to obtain the Hessian elements.

\def\cpl{Chem. Phys. Lett.}

\newpage

\begin{table}
\caption{Parameters of the interaction potential~\cite{sylvie-94,charmm}.}
\begin{center}
\begin{tabular}{|c|c|c|c|c|c|c|c|}
\hline
Stretch & \multicolumn{3}{c|}{$r_{0}$ [\AA]} & \multicolumn{4}{c|}{$k_{\rm r}$ [K \AA$^{-2}$]} \\ \hline
 C-C & \multicolumn{3}{c|}{1.53} & \multicolumn{4}{c|}{237017.0} \\
 C-O & \multicolumn{3}{c|}{1.43} & \multicolumn{4}{c|}{171094.8} \\
 C-H & \multicolumn{3}{c|}{1.09} & \multicolumn{4}{c|}{Constrained} \\
 H-H & \multicolumn{3}{c|}{1.78} & \multicolumn{4}{c|}{Constrained} \\
\hline
\multicolumn{8}{|c|}{} \\
\hline
Bend & \multicolumn{3}{c|}{$\theta_{0}$} & \multicolumn{4}{c|}{$k_{\theta}$ [K rad$^{-2}$]} \\
\hline
C-O-C &\multicolumn{3}{c|}{112$^{o}$}  &\multicolumn{4}{c|}{110255.50}\\
O-C-C &\multicolumn{3}{c|}{110$^{o}$}  &\multicolumn{4}{c|}{76942.34}\\
O-C-H &\multicolumn{3}{c|}{109.5$^{o}$}  &\multicolumn{4}{c|}{30193.20}\\
H-C-C &\multicolumn{3}{c|}{110$^{o}$}  &\multicolumn{4}{c|}{45199.50}\\
\hline
\multicolumn{8}{|c|}{} \\
\hline
Torsions & $a_0$ [K] & $a_1$ [K] & $a_2$ [K]& $a_3$ [K] & $a_4$ [K]& $a_5$ [K]& 
            $a_6$ [K] \\
\hline
C-C-O-C & -50.322 & 150.966 & 0.0 & -201.288 & 0.0 & 0.0 & 0.0 \\
C-O-C-H & -50.322 & 150.966 & 0.0 & -201.288 & 0.0 & 0.0 & 0.0 \\
H-C-C-H &  83.03  &-249.090 & 0.0 &  332.120 & 0.0 & 0.0 & 0.0 \\
O-C-C-O &  265.70 &-1826.19 & 2144.72 & 3901.46 & -1667.17 & 142.91 & 1480.98\\
H-C-C-O &  98.128 &-294.384 & 0.0 & 392.512 & 0.0 & 0.0 & 0.0 \\
\hline
\multicolumn{8}{|c|}{} \\
\hline
Non-bonded & \multicolumn{3}{c|}{A [K]} & B [$\AA^{-1}$]  &\multicolumn{3}{c|}{C [K$\AA^{6}$]} \\
\hline
C...C & \multicolumn{3}{c|}{15909350.62} & 3.3058 & \multicolumn{3}{c|}{325985.916}\\
C...O & \multicolumn{3}{c|}{21604039.75} & 3.6298 & \multicolumn{3}{c|}{177536.016}\\
C...H & \multicolumn{3}{c|}{7571800.37}  & 3.6832 & \multicolumn{3}{c|}{91334.430} \\
O...O & \multicolumn{3}{c|}{29337172.46} & 4.0241 & \multicolumn{3}{c|}{96668.562} \\
O...H & \multicolumn{3}{c|}{10282092.97} & 4.0900 & \multicolumn{3}{c|}{49718.136} \\ 
H...H & \multicolumn{3}{c|}{3603659.064} & 4.1580 & \multicolumn{3}{c|}{25563.576} \\ 
\hline
\multicolumn{8}{|c|}{} \\
\hline
\multicolumn{8}{|c|}{Atomic charges [e]} \\
\hline
\multicolumn{4}{|c|}{q$_{\rm C}$} & \multicolumn{4}{c|}{0.103} \\
\multicolumn{4}{|c|}{q$_{\rm O}$} & \multicolumn{4}{c|}{-0.348} \\
\multicolumn{4}{|c|}{q$_{\rm H}$} & \multicolumn{4}{c|}{0.0355} \\
\hline
\end{tabular}
\end{center}
\end{table}

\begin{table}
\caption{Lattice parameters obtained from simulation compared to experiment~\cite{takahashi}.}
\begin{center}
\begin{tabular}{|c|c|c|}\hline
Lattice parameter & Simulation & Experiment \\ \hline
a $[\AA]$ & 8.08 & 8.05 \\
 & & \\
b $[\AA]$ & 13.17 & 13.04 \\
 & & \\
c $[\AA]$ & 18.45 & 19.48 \\
 & & \\
$\alpha [^{o}]$ & 89.98 & 90.0 \\ 
 & & \\
$\beta [^{o}]$ & 123.01 & 125.40 \\ 
 & & \\
$\gamma [^{o}]$ & 89.99 & 90.0 \\ \hline
\end{tabular}
\end{center}
\end{table}
\clearpage

\newpage
\begin{subfigures}
\begin{figure}
\centerline{\psfig{figure=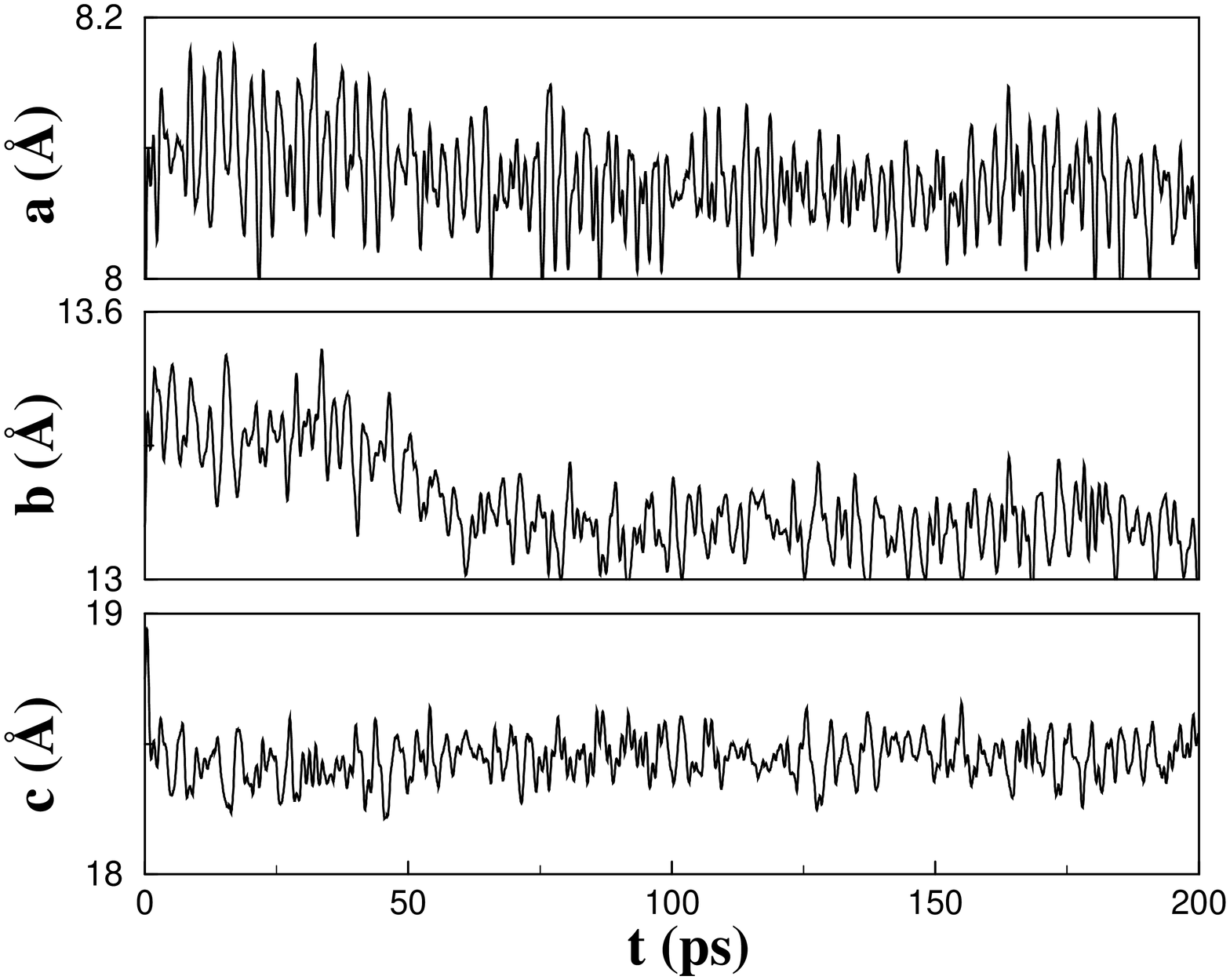,height=5.5in,angle=270}}
\vspace*{1.0cm}
\caption{}
\end{figure}

\newpage

\begin{figure}
\centerline{\psfig{figure=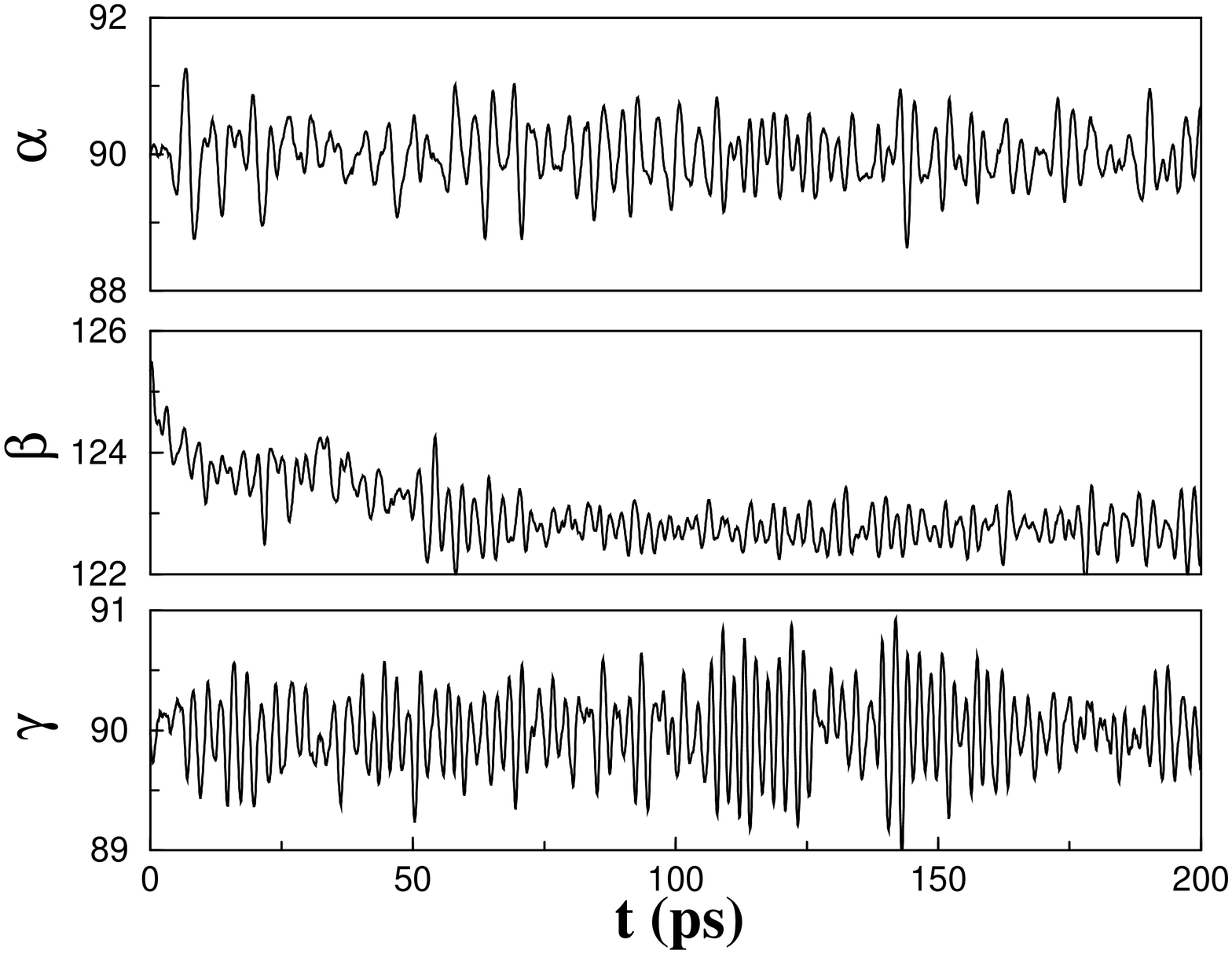,height=5.5in,angle=270}}
\vspace*{1.0cm}
\caption{}
\vspace*{1.0cm}
{\bf Figure 1:}~~~Plot of instantaneous cell parameters of PEO crystal obtained from simulations :
(a) cell lengths, (b) cell angles.  The zero time configuration corresponds to
 the experimental crystal structure.
\end{figure}
\end{subfigures}

\newpage
\begin{subfigures}
\begin{figure}
\centerline{\psfig{figure=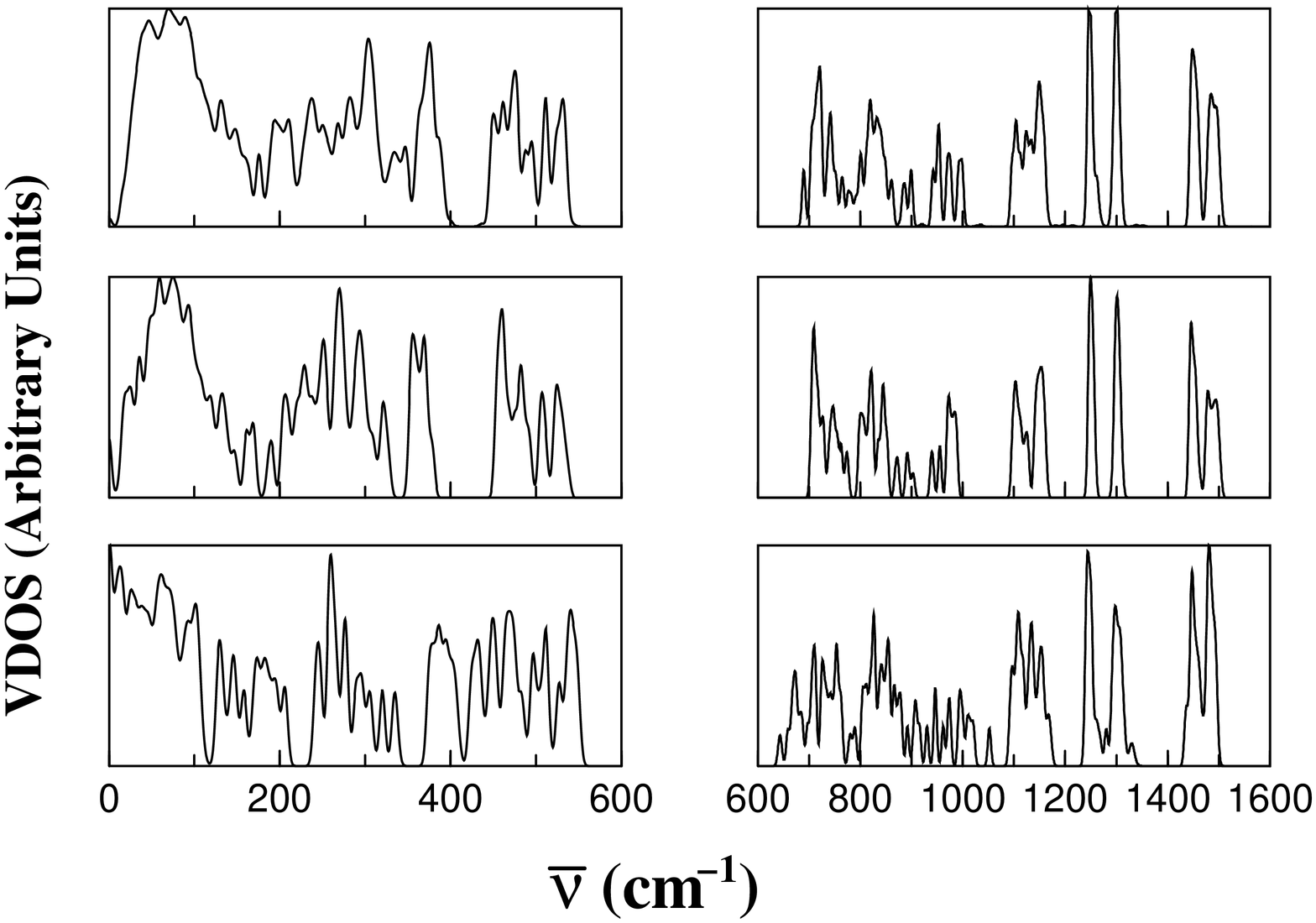,height=5.2in,angle=270}}
\vspace{1.0cm}
\caption{}
\end{figure}

\newpage

\begin{figure}
\centerline{\psfig{figure=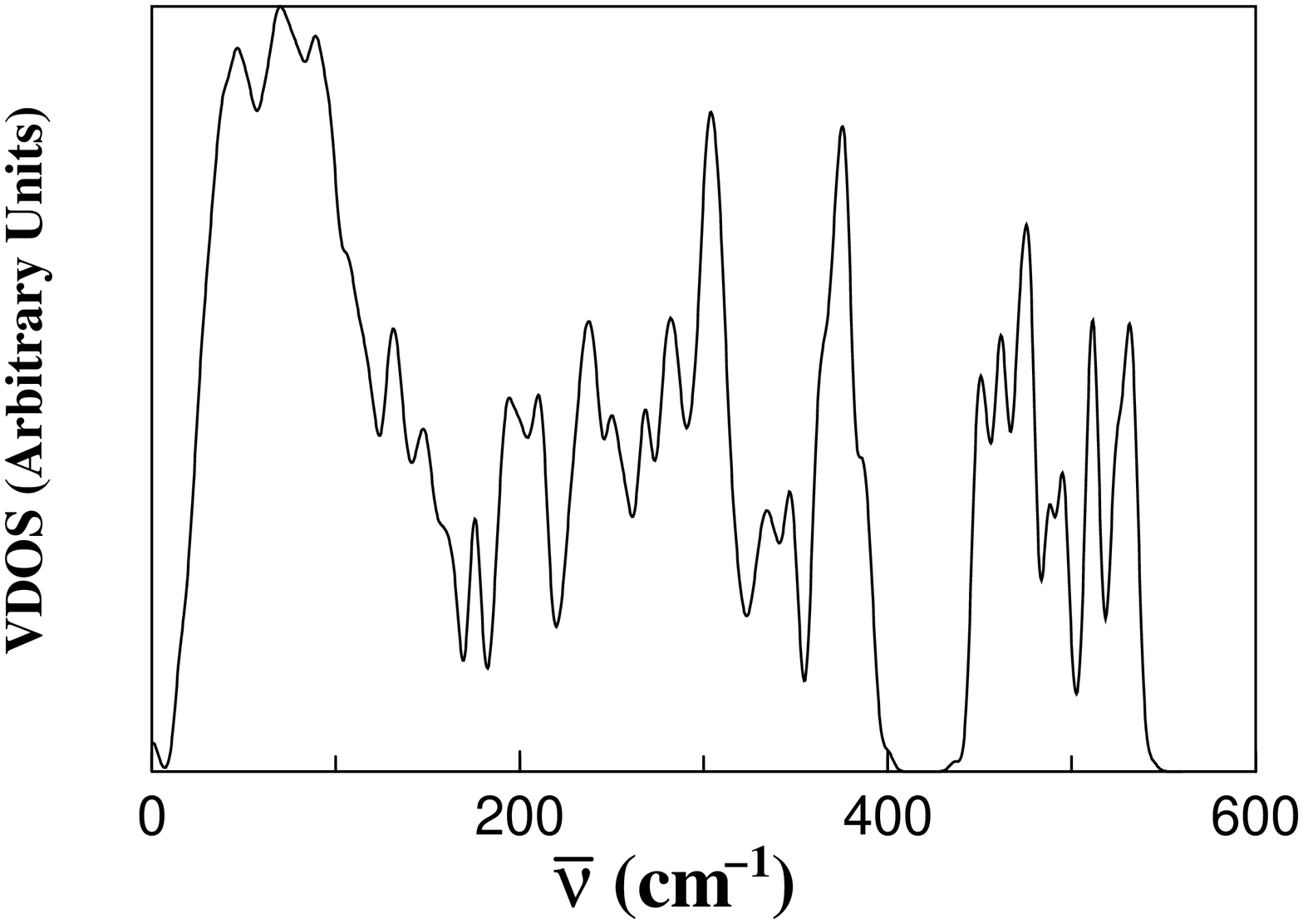,height=5.0in,angle=270}}
\vspace{1.0cm}
\caption{}
\end{figure}

\newpage
\begin{figure}
\centerline{\psfig{figure=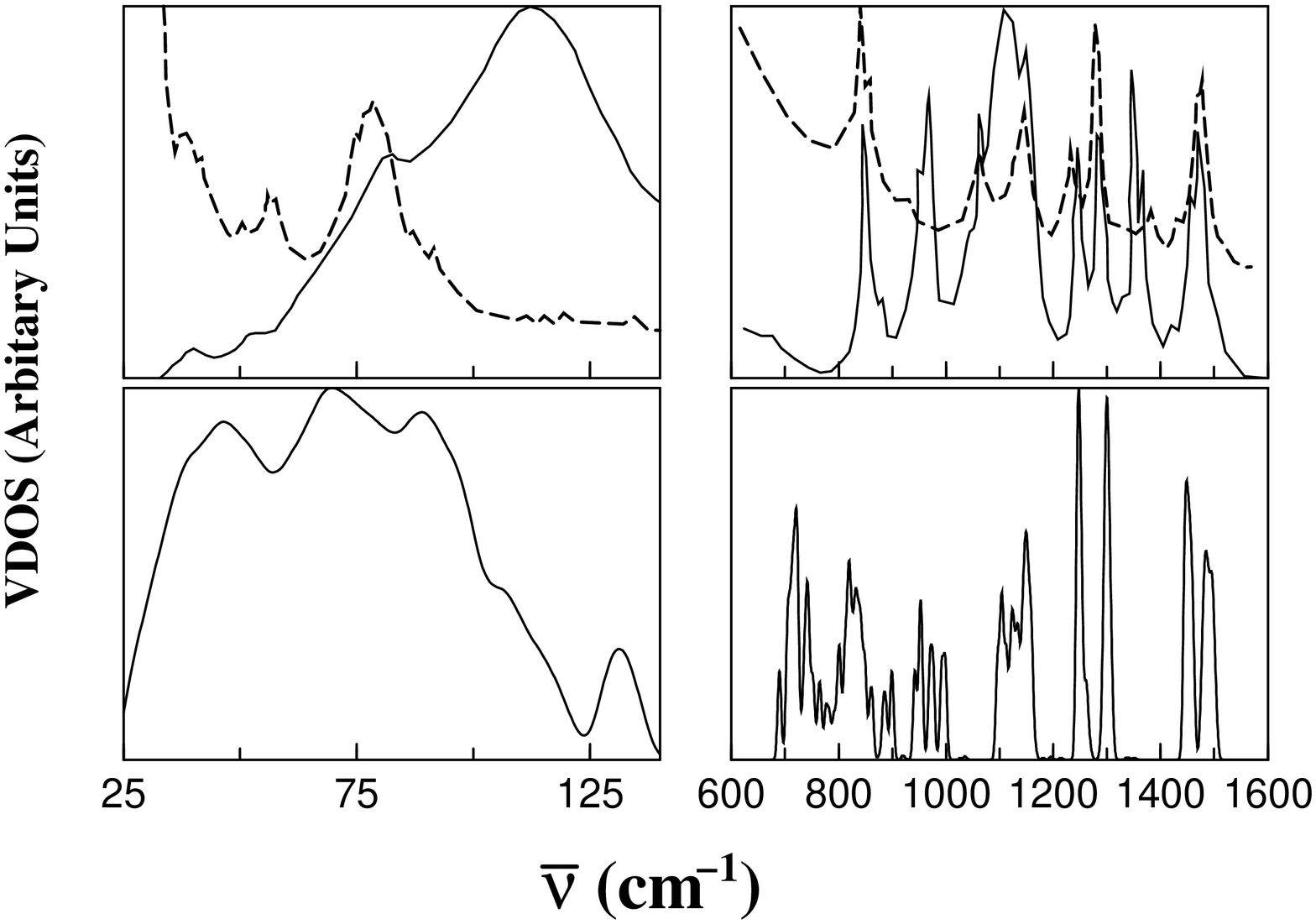,height=5.0in,angle=270}}
\caption{}

\vspace*{0.5cm}
{\bf Figure 2:}~~~(a) The vibrational density of states of a single molecule (bottom), for one unitcell (middle) and for the
 $\langle422\rangle$ crystal (top) of PEO, each obtained from normal mode analysis.
The two sections of each
spectrum are normalized independently to enable better comparison among the three system sizes.
The spectra in Figures~2a, 2b, and 2c are convoluted with a Gaussian function of width 4cm$^{-1}$.
(b) The vibrational density of states of the $\langle422\rangle$ crystal, obtained by NMA shown in expanded scale.
(c) Vibrational density of states obtained from the NMA method (bottom panels)
are compared with experimental Raman (Dashed lines) and Infra-red (Continuous lines)
absorption intensities.
Experimental data presented in the top left and top right panels
were obtained from Ref.~\cite{rabolt} and Ref.~\cite{yoshihara} respectively.
Note that the simulated spectrum is the raw
density of states and does not contain any other terms that are needed to calculate the
experimental absorption spectra. Thus only the peak positions are comparable, and not their intensities.
\end{figure}
\end{subfigures}

\newpage
\begin{figure}
\centerline{\psfig{figure=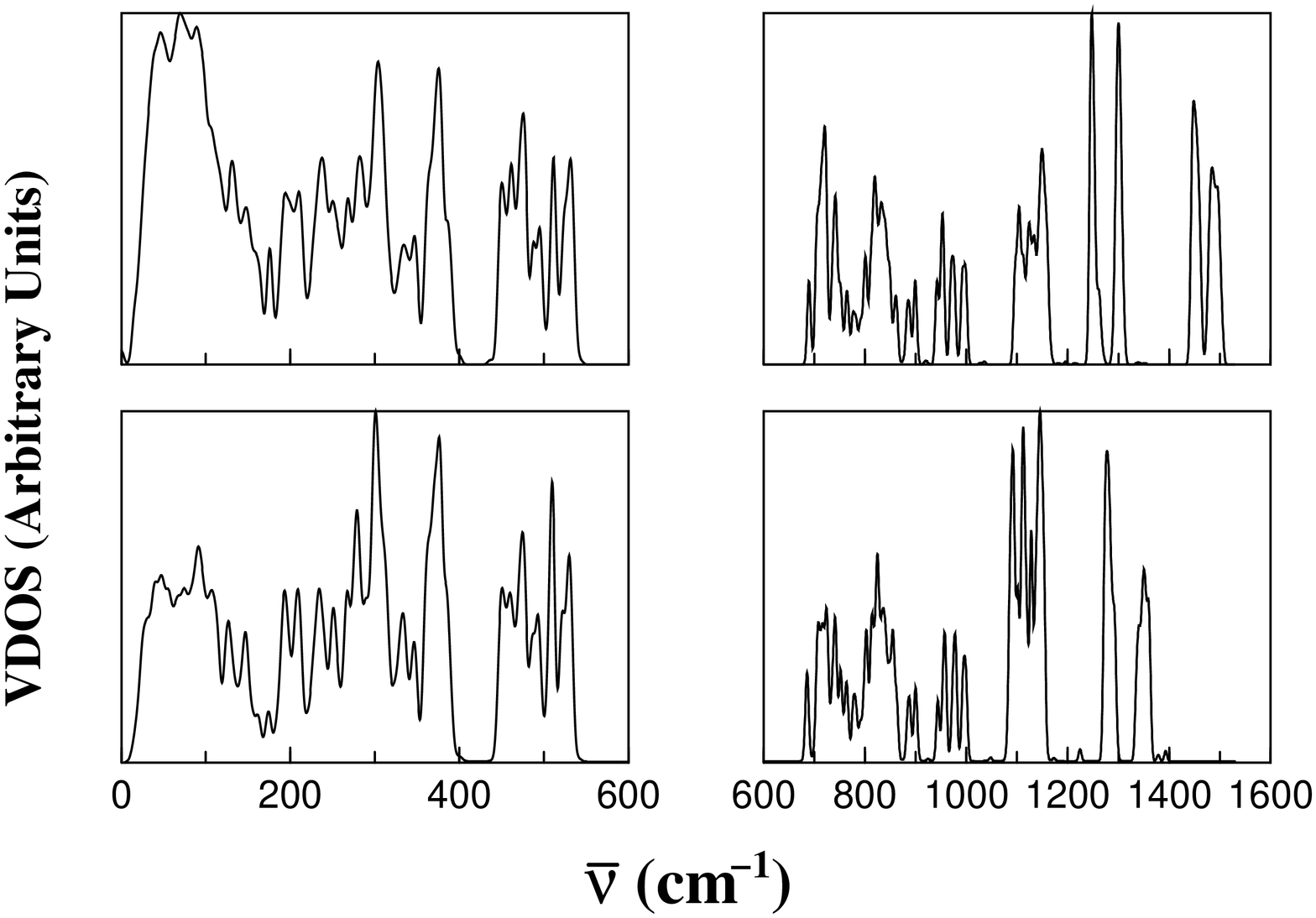,height=4.5in,angle=270}}
\vspace*{2.0cm}
\caption{
The vibrational density of states of $\langle422\rangle$ PEO crystal obtained from normal mode
analysis (top panels) is compared with that obtained as the power spectrum of the
velocity autocorrelation function of all atoms (bottom panels).
The two sections of each spectrum are normalized independently to enable better
comparison among the two methods.
The spectra are convoluted with a Gaussian function of width 4cm$^{-1}$.}
\end{figure}

\newpage
\begin{subfigures}
\begin{figure}
\centerline{\psfig{figure=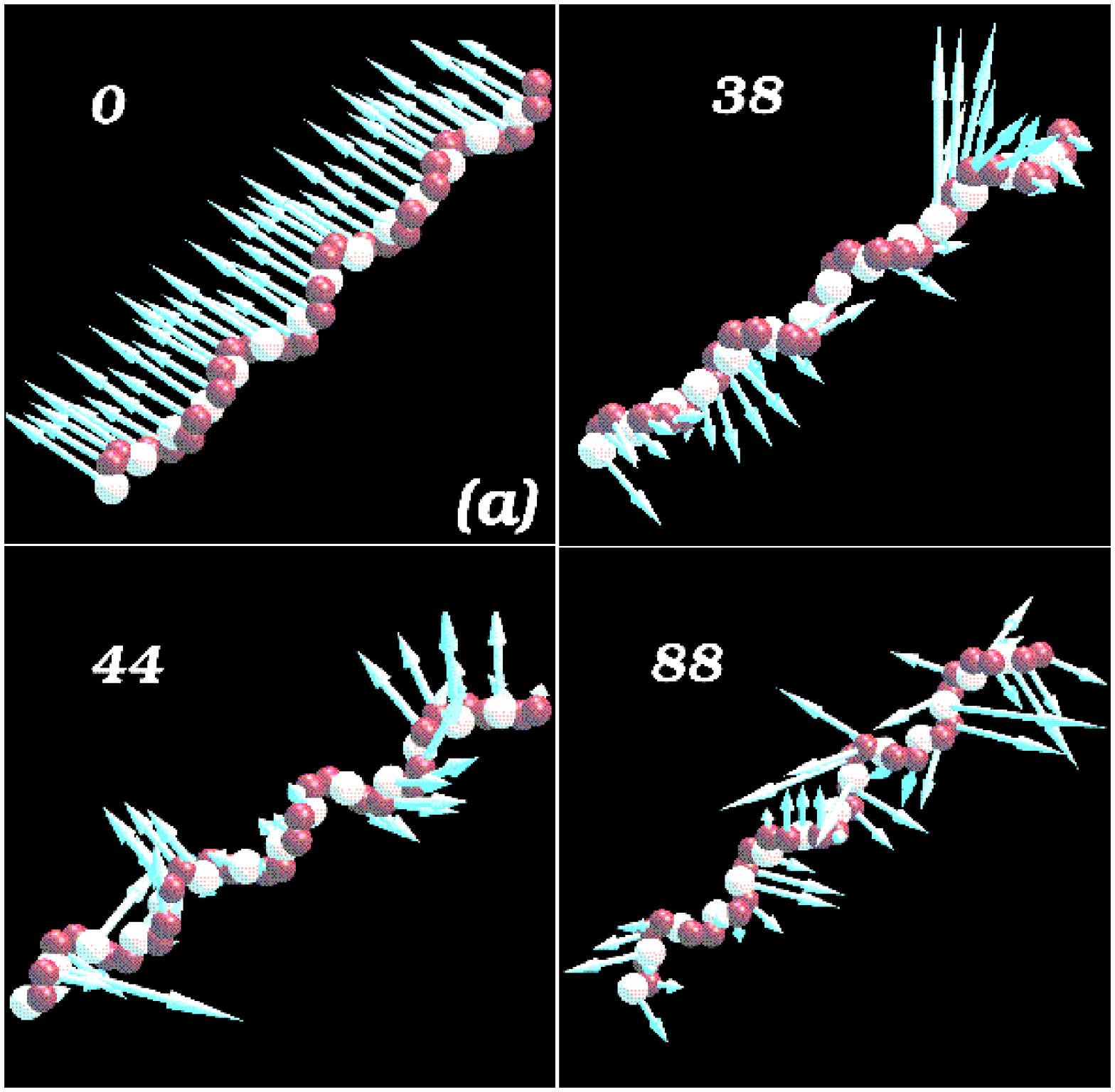,height=5.0in,angle=270}}
\vspace{0.5cm}
\caption{}
\end{figure}

\newpage
\begin{figure}
\centerline{\psfig{figure=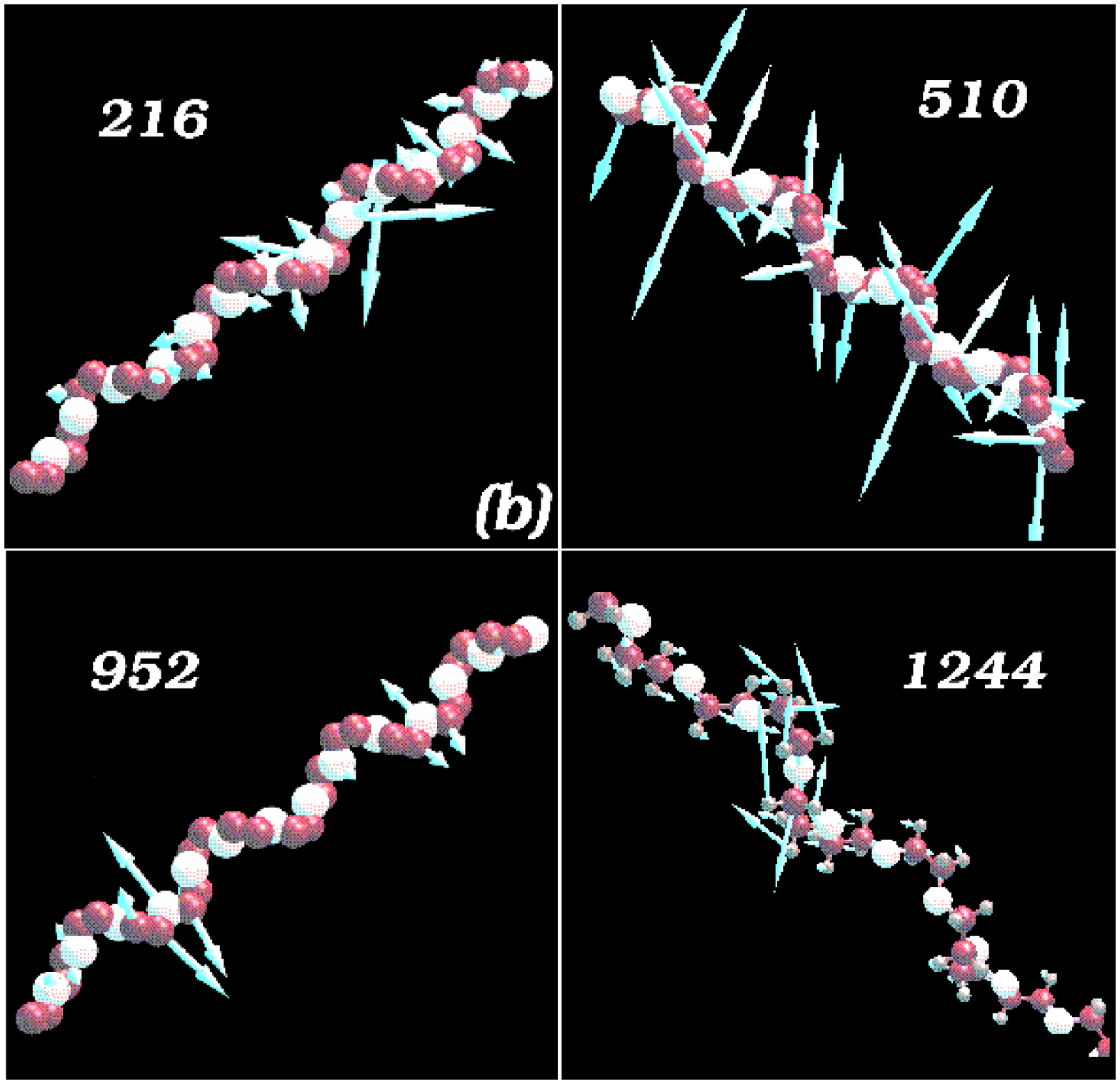,height=5.0in,angle=270}}
\vspace*{0.5cm}
\caption{}
\vspace*{0.5cm}

{\bf Figure 4:}~~~~(a),(b) : Atomic displacements of representative normal modes of the PEO crystal.
For each mode, one of the chains which exhibit
significant atomic displacement is shown, for clarity. Red (or Black) spheres denote carbon atoms and
white spheres denote oxygens. Hydrogen atoms are not shown for clarity in all the modes, except the
one with frequency 1244~cm$^{-1}$. Arrows denote the directions of atomic displacements and their
lengths are scaled up for the purpose of visualization. The mode
at 0~cm$^{-1}$ is the rigid body translation of the chain, and that at
1244~cm$^{-1}$ arises from wagging of CH$_2$ groups. Frequencies in cm$^{-1}$ are as
provided in the figures. The representation of these modes in the D$_7$ group, are as
follows, with frequencies in cm$^{-1}$ given in paranthesis: A$_2$ (38), A$_2$ (44),
E (88), A$_1$ (216), E (510), E (952), and E (1244).
\end{figure}
\end{subfigures}

\newpage
\begin{figure}
\centerline{\psfig{figure=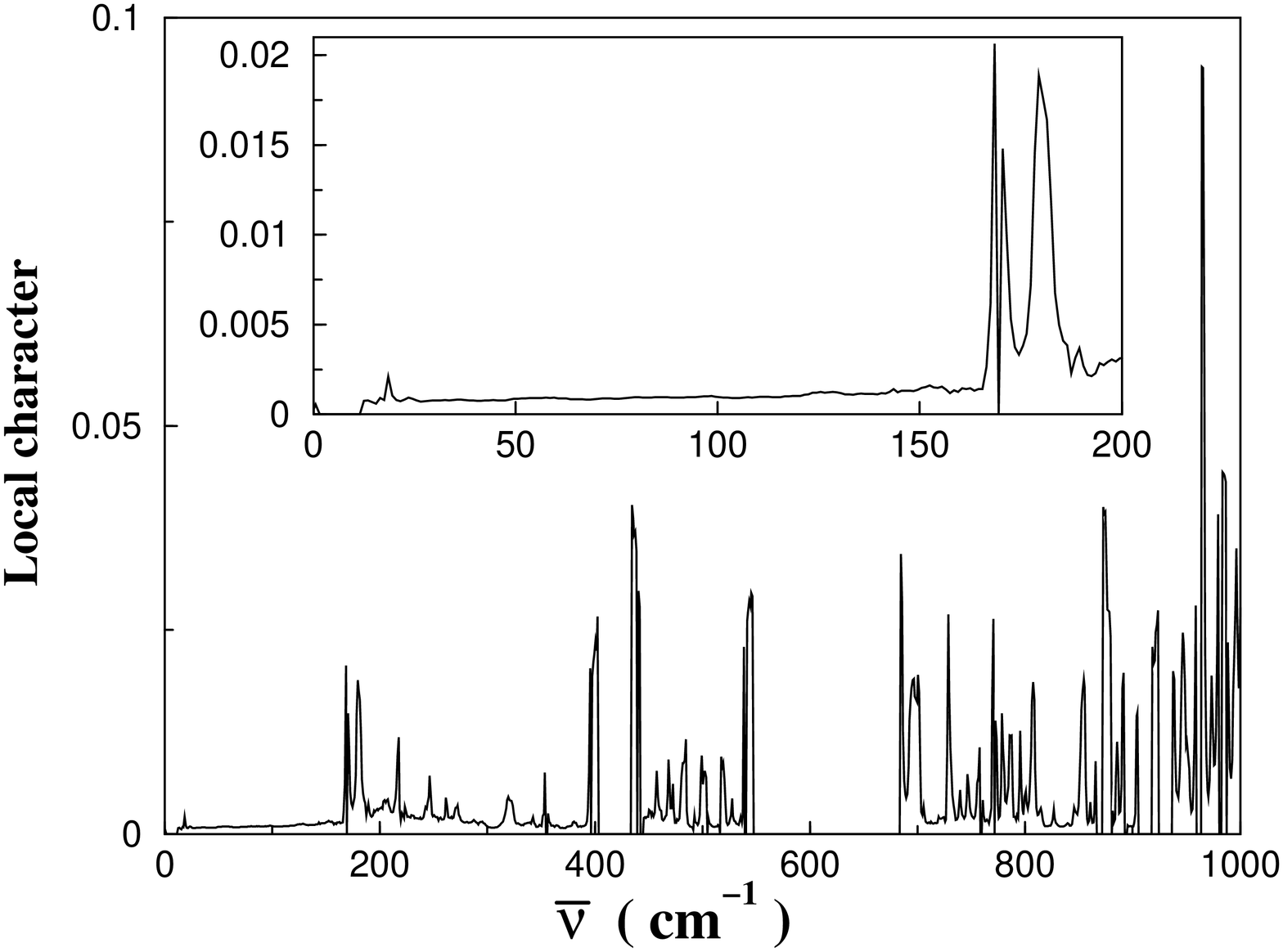,height=4.5in,angle=270}}
\vspace*{2.0cm}
\caption{
Local character indicator for the normal modes of PEO crystal. The inset
shows the frequency range where
 collective motion is most probable with very small local character indicator values.}
\end{figure}

\newpage
\begin{figure}
\centerline{\psfig{figure=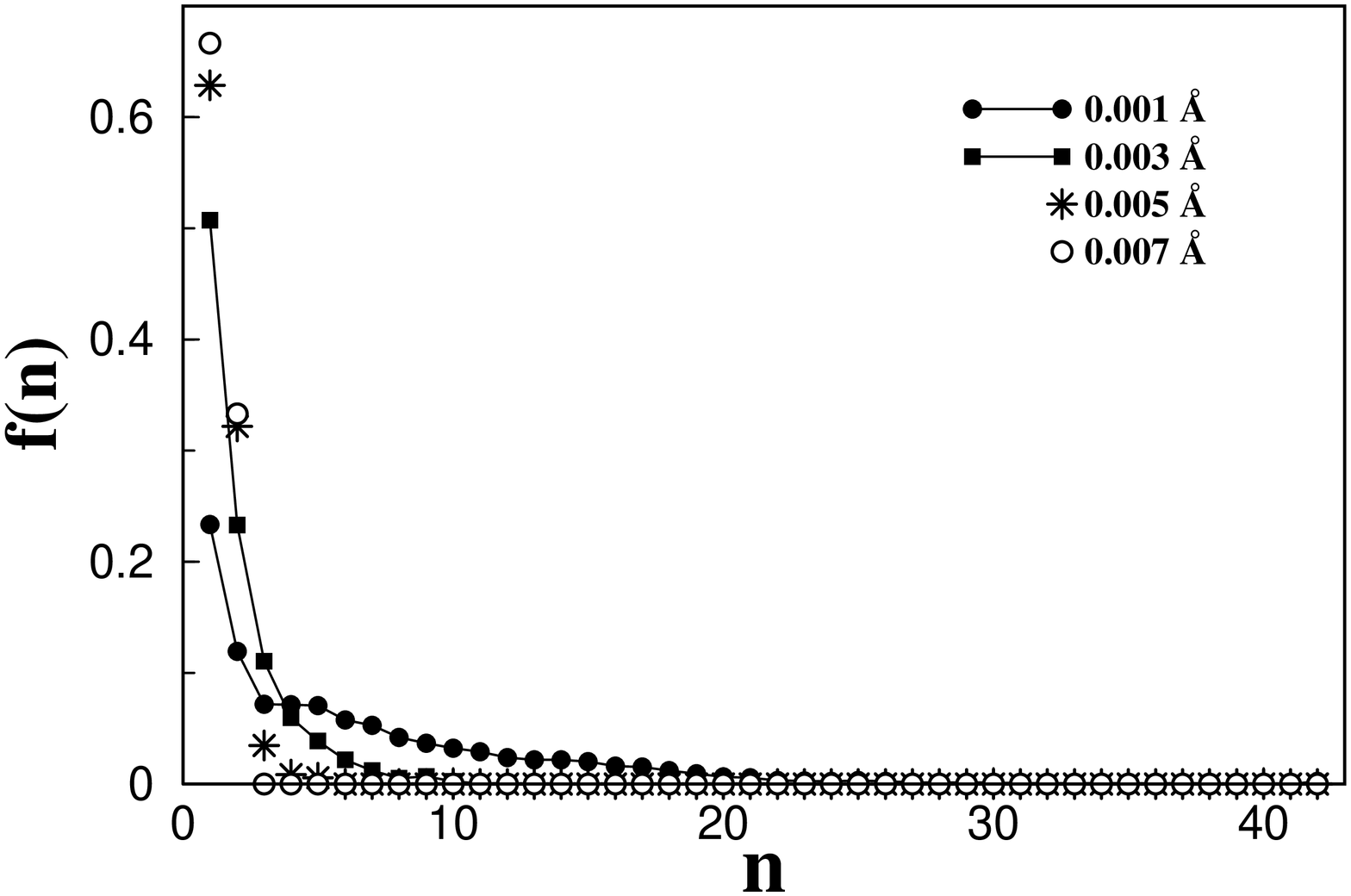,height=4.5in,angle=270}}
\vspace*{2.0cm}
\caption{
The segment size distribution, f(n), corresponding to the mode at 44~cm$^{-1}$ of the PEO crystal.
The distribution is shown for four different displacement cutoffs.
Such distributions are used to calculate  the average continuous segment size ($<$CSS$>$) which
is defined in equation (7).}
\end{figure}

\newpage
\begin{figure}
\centerline{\psfig{figure=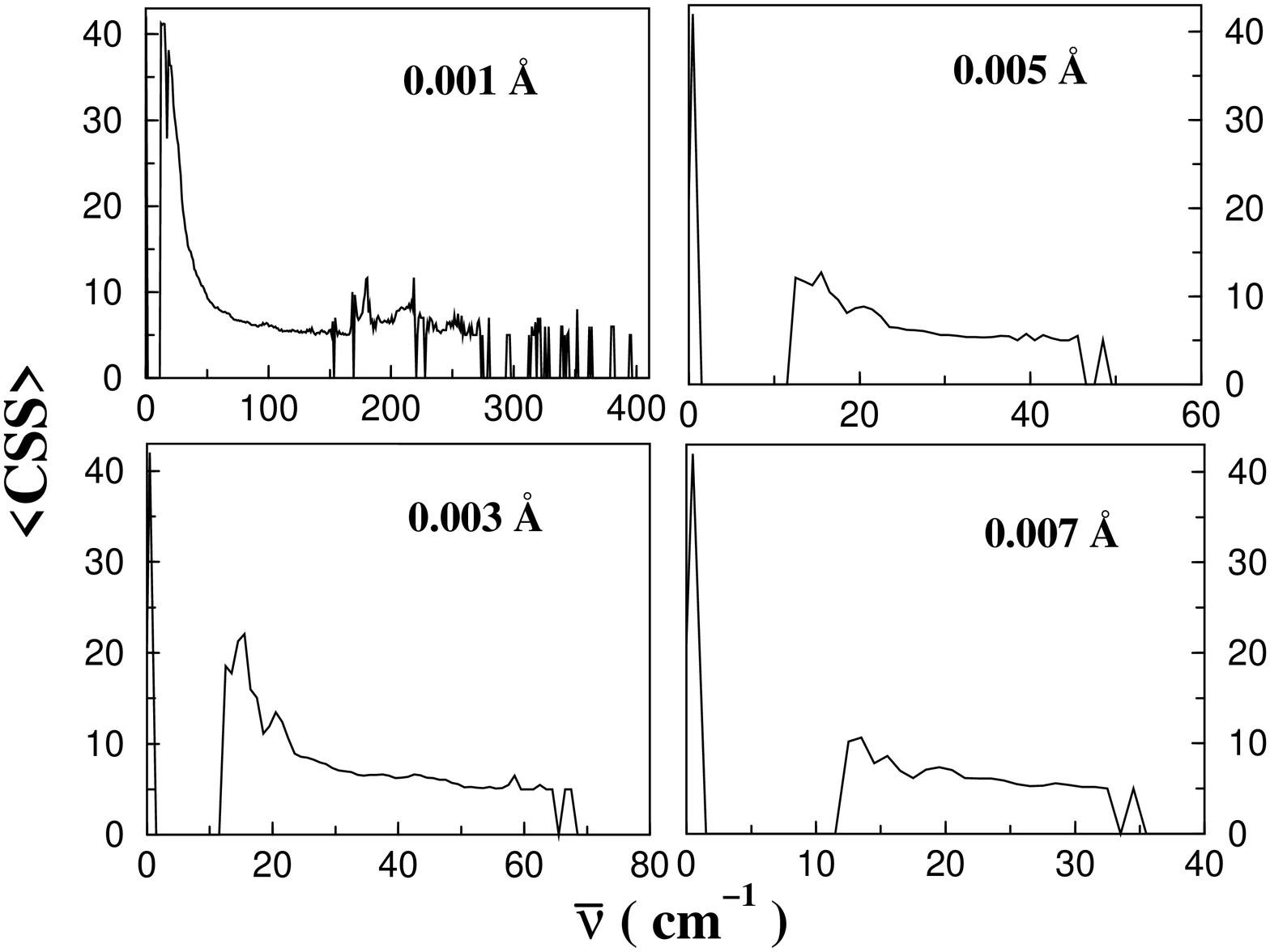,height=4.5in,angle=270}}
\vspace*{2.0cm}
\caption{
Variation of the average continuous segment size ($<$CSS$>$) with respect to the
vibrational frequencies for different displacement cutoffs.}
\end{figure}


\newpage
\begin{thebibliography}{99}
\bibitem{halley_power} J.W. Halley, and Y. Duan, J. Power Sources {\bf 89}, 139 (2000).
\bibitem{wakihara} M. Wakihara, Mat. Sci. Engg. {\bf R33}, 109 (2001).
\bibitem{brucebook} P.G. Bruce, {\em Solid State Electrochemistry}, Cambridge University Press, Cambridge, 1995.
\bibitem{bruce1} P.G. Bruce, and C.A. Vincent, J. Chem. Soc. Faraday Trans. {\bf 89}, 3187 (1993).
\bibitem{angell_jpcm} X.G. Sun, W. Xu, S.S. Zhang, and C.A. Angell, J. Phys. Cond. Matt. {\bf 13},
 8235 (2001). 
\bibitem{donoso} J.P. Donoso, T.J. Bonagamba, H.C. Panepucci, L.N. Oliveira, W. Gorecki, C. Berthier and M. Armand, J. Chem. Phys. {\bf 98}, 10026 (1993).
\bibitem{boden} I.M. Ward, N. Boden, J. Cruickshank, and S.A. Leng, 
Electrochim. Acta. {\bf 40}, 2071 (1995).
\bibitem{qens} G. Mao, R.F. Perea, W.S. Howells, D.L. Price, and M.-L. Saboungi, Nature 
(London) {\bf 405}, 163 (2000); 
M.-L. Saboungi, D.L. Price, G. Mao, R.F. Perea, O. Borodin, and G.D. Smith, 
M. Armand, and W.S. Howells, Solid State Ionics {\bf 147}, 225 (2002);
\bibitem{angell} C.A. Angell, Solid State Ionics {\bf 9-10}, 3 (1983).
\bibitem{berthier} C. Berthier, W. Gorecki, M. Minier, M.B. Armand, J.M. Chanbagno, 
and P. Rigaud, Solid State Ionics {\bf 11}, 91 (1983).
\bibitem{gorecki} W. Gorecki, P. Donoso, C. Berthier, M. Mali, J. Roos, 
D. Brinkmann, M.B. Armand, Solid State Ionics {\bf 28-30},1018 (1988).
\bibitem{bruce2} Z. Gadjourova, Y.G. Andreev, D.P. Tunstall, and P.G. Bruce, Nature {\bf 412}, 520 (2001)
\bibitem{leeuw} B. Mos, P. Verkerk, A. van Zon, and S.W. de Leeuw, Physica B {\bf 276}, 351 (2000);
 S.W. de Leeuw, A. Van Zon, and G.J. Bel, Electrochim. Acta {\bf 46}, 1419 (2001);
 J.J. de Jonge, A. van Zon, and S.W. de Leeuw, Solid State Ionics {\bf 147}, 349 (2002).
\bibitem{sylvie-94} S. Neyertz, D. Brown, and J.O. Thomas, J. Chem. Phys. {\bf 101}, 10064 (1994).
\bibitem{thesis} S. Neyertz, Ph.D. thesis, Uppsala University, Uppsala, 1995. 
\bibitem{neyertzmelt} S. Neyertz, and D. Brown, J. Chem. Phys. {\bf 102}, 9725 (1995).
\bibitem{sylvie-acta95} S. Neyertz, D. Brown, and J.O. Thomas, Electrochim. Acta  {\bf 40}, 2063 (1995).
\bibitem{sylvie-cps95} S. Neyertz, D. Brown, and J.O. Thomas, Comput. Polym. Sci.  {\bf 5}, 107 (1995).
\bibitem{florian-95} F. Muller-Plathe, and W.F. van Gunsteren, J. Chem. Phys. {\bf 103}, 4745 (1995).
\bibitem{laasonen} K. Laasonen, and M.L. Klein, J. Chem. Soc. Faraday Trans. {\bf 91}, 2633 (1995).
\bibitem{halley2001} J.W. Halley, Y. Duan, B. Nielsen, P.C. Redfern, and L.A. Curtiss, 
J. Chem. Phys. {\bf 115}, 3957 (2001); B. Lin, P.T. Boinske, and J.W. Halley, J. Chem. Phys. 
{\bf 105}, 1668 (1996).
\bibitem{smith_peo} P. Ahlstr$\ddot{o}$m, G. Wahnstr$\ddot{o}$m, P. Carlsson, O. Borodin,
and G.D. Smith, J. Chem. Phys. {\bf 112}, 10669 (2000).
\bibitem{smith} O. Borodin, and G.D. Smith, Macromolecules {\bf 31}, 8396 (1998); Macromolecules {\bf 33}, 2273 (2000);
\bibitem{smith_h2o} O. Borodin, F. Trouw, D. Bedrov, and G.D. Smith, J. Phys. Chem. B {\bf 106}, 5184 (2002);
O. Borodin, D. Bedrov, and G.D. Smith, J. Phys. Chem. B {\bf 106}, 5194 (2002);
 G.D. Smith, D. Bedrov, and O. Borodin, Phys. Rev. Lett. {\bf 85}, 5583 (2000).
\bibitem{smith_spectrochim} G.D. Smith, O. Borodin, M. Pekny, B. Annis, D. Londono, and R.L. Jaffe,
Spectrochim. Acta A {\bf 53}, 1273 (1997).
\bibitem{carini} G. Carini, G. D'Angelo, G. Tripodo, A. Bartolotta, and G. Di Marco
    Phys. Rev. B {\bf 54}, 15056-15063 (1996). 
\bibitem{mustarelli}  P. Mustarelli, C. Capiglia, E. Quartarone, C. Tomasi, 
P. Ferloni, and L. Linati, Phys. Rev. B {\bf 60}, 7228 (1999).
\bibitem{elliott_papers} S.I. Simdyankin, M. Dzugutov, S.N. Taraskin, and S.R. Elliott, 
Phys. Rev. B {\bf 63}, 184301 (2001); S.N. Taraskin, and S.R. Elliott, Phys. Rev. B 
{\bf 61}, 12017 (2000); S.N. Taraskin, and S.R. Elliott, Phys. Rev. B {\bf 59}, 8572 (1999).
\bibitem{nagel} C.S. O'Hern, S.A. Langer, A.J. Liu, and S.R. Nagel, Phys. Rev. Lett. {\bf 86}, 
111 (2001).
\bibitem{hess_avg}  K. Fukui, B.G. Sumpter, D.W. Noid, C. Yang, and R.E. Tuzun,
   J. Phys. Chem. B, {\bf 104}, 526 (2000); D.W. Noid, K. Fukui, B.G. Sumpter, C. Yang, and R.E. Tuzun, \cpl {\bf 316}, 285 (2000). 
\bibitem{karplus} H.W.T. van Vlijmen and M. Karplus, J. Phys. Chem. B {\bf 103}, 3009 (1999).
\bibitem{verma_jpcb} N.P. Barton, C.S. Verma, L.S.D. Caves, J. Phys. Chem. B {\bf 107}, 2170 (2003); 
J. Phys. Chem. B {\bf 106}, 11036 (2002).
\bibitem{takahashi} Y. Takahashi and H. Tadokoro, Macromolecules {\bf 6}, 672 (1973).
\bibitem{martyna} G.J. Martyna, M.L. Klein, and M. Tuckerman, J. Chem. Phys. {\bf 97}, 2635 (1992).
\bibitem{piny-md} M.E. Tuckerman, D.A. Yarne, S.O. Samuelson, A.L. Hughes, 
and G. Martyna, Comput. Phys. Commun. {\bf 128}, 333 (2000).
\bibitem{hansen} J.-P. Hansen, in {\em Molecular Dynamics Simulations of Statistical Mechanics Systems},
Ed. G. Ciccotti, and W.G. Hoover, (North-Holland, Amsterdam, 1986).
\bibitem{charmm} A.D. MacKerell Jr. {\em et al}, J. Phys. Chem. B 
{\bf 102}, 3586 (1998).
\bibitem{abramovitz} M. Abramowitz, and I.A. Stegun, {\em Handbook of 
Mathematical Functions}, (Dover, New York, 1970).
\bibitem{elliott} S.N. Taraskin, and S.R. Elliott, Phys. Rev. B. {\bf 56}, 8605 (1997).
\bibitem{parr-rahman} M. Parrinello, and A. Rahman, Phys. Rev. Lett. {\bf 45}, 1196 (1980).
\bibitem{rabolt} J.F. Rabolt, K.W. Johnson, and R.N. Zitter, J. Chem. Phys. {\bf 61}, 504, (1974).
\bibitem{dacosta} V.M. Da Costa, T.G. Fiske, and L.B. Coleman, J. Chem. Phys. {\bf 101}, 2746, (1994).
\bibitem{branca} C. Branca, A. Faraone, S. Magazu, G. Maisano, P. Migliardo, and V. Villari, 
J. Mol. Liquids {\bf 87}, 21 (2000).
\bibitem{yoshihara} T. Yoshihara, H. Tadokoro, S. Murahashi, J. Chem. Phys. {\bf 41}, 2902 (1964).
\bibitem{lam} K. Song, and S. Krimm, J. Polym. Sci. Polym. Phys. Ed. {\bf 28}, 35, (1990).
\bibitem{case} D.A. Case, Curr. Opin. Struct. Biol. {\bf 4}, 285 (1994).
\end{thebibliography}
\end{document}